\documentclass[11pt,a4paper]{article}
\usepackage{amsmath,bm,graphicx,booktabs}
\usepackage{geometry}
\usepackage{indentfirst,float,appendix,lmodern}
\usepackage{multirow}
\usepackage[authoryear]{natbib}
\usepackage{algorithm,algorithmic}
\usepackage[colorlinks,citecolor=blue,urlcolor=blue]{hyperref}
\usepackage{inputenc}
\usepackage[english]{babel}
\usepackage{amsthm}
\usepackage{color}
\usepackage{courier}
\usepackage{amssymb}
\usepackage{mathrsfs}
\usepackage{enumitem}
\usepackage{caption}
\usepackage{comment}
\graphicspath{{figures/}}

\providecommand{\keywords}[1]{\textbf{\textit{key words: }} #1}

\newtheorem{remark}{Remark}

\newcommand{\rc}{\color{black}}

\title{Clustering of longitudinal curves via a penalized method and EM algorithm}

\date{}

\author{ Xin Wang \thanks{Email: xwang14@sdsu.edu}\\
	Department of Mathematics and Statistics, San Diego State University}

\begin{document}

\maketitle

\begin{abstract}
In this article, a new method is proposed for clustering longitudinal curves.  In the proposed method, clusters of mean functions are identified through a weighted concave pairwise fusion method. The EM algorithm and the alternating direction method of multipliers algorithm are combined to estimate the group structure, mean functions and principal components simultaneously.  The proposed method  also allows to incorporate the prior neighborhood information to have more meaningful groups by adding pairwise weights in the pairwise penalties. In the simulation study, the performance of the proposed method is compared to some existing clustering methods in terms of the accuracy for estimating the number of subgroups and mean functions. The results suggest that ignoring the covariance structure will have a great effect on the performance of estimating the number of groups and estimating accuracy. The effect of including pairwise weights is also explored in a spatial lattice setting to take into consideration of the spatial information. The results show that incorporating spatial weights will improve the performance. A real example is used to illustrate the proposed method.
\end{abstract}

\keywords{ADMM algorithm, B-spline regression,  Clustering, EM algorithm, Functional principal component analysis, Penalty functions}

\section{Introduction}
\label{sec_intro}

Clustering is a method to identify homogeneous subgroups from a heterogeneous population. \cite{jain2010data} had a review of different clustering methods. One particular type of clustering problems is to find clusters for longitudinal curves. There are a lot of applications in clustering longitudinal curves, such as bioscience \citep{zhu2019clustering}, bioinformatics \citep{ng2006mixture}, geostatistics \citep{chiou2008correlation} and social science \citep{jiang2012clustering}.
As mentioned in \cite{zhu2018cluster}, traditional clustering methods don't take time ordering into account. Besides time ordering, traditional clustering methods don't consider covariance structure. To solve this problem, different clustering methods are developed. 

If we consider that longitudinal observations are from some functions over time,  we can use the framework of functional data to analyze longitudinal data \citep{ramsay2005fda}.  In functional data analysis,  longitudinal curves are assumed to be functions of time,  but functions are only observed on discrete time points.  \cite{jacques2014functional} provided an overview of some functional clustering methods. \cite{james2003clustering} proposed a model-based method for clustering sparse sampled functional data, where spline basis was used to model mean curves. \cite{peng2008distance} considered a distance-based clustering approach that defined the distance between two functions. \cite{luan2003clustering} and \cite{coffey2014clustering} both used mixed effects models to find clusters in time-course gene expression data. Some related works are based on functional principal components analysis (FPCA). In functional data analysis, FPCA is a useful tool to model mean curves and covariance functions \citep{yao2005functional, li2010uniform}.  \cite{chiou2007functional} and \cite{chiou2008correlation} proposed functional clustering methods based on principal components and $k$-means. \cite{sangalli2010k} (KMA) developed an algorithm to cluster and align curves jointly.  \cite{bouveyron2011model} (funHDDC) built a procedure  based on a functional latent mixture model for clustering functional data.  A method based on functional mixture models and discriminative functional  subspace was proposed in  \cite{bouveyron2015discriminative} (FEM) to find clusters of curves. \cite{jacques2013funclust}  defined an approach using a mixture model when assuming a Gaussian distribution of the principal components.  Their approach was based on an approximation of the notion of probability density for functional random variables and Karhunen-Lo{\`e}ve expansion.

These methods mentioned above cannot incorporate extra information,  such as locations.  In the traditional clustering problem, constrained clustering is discussed to use extra information or labeled data, such as \cite{basu2004active} and \cite{de2012constrained}. Must-link and cannot-link are needed. Instead of defining two sets of links, \cite{chi2015splitting} considered all pairwise links and constructed an optimization problem for clustering based on pairwise $L_p (p\geq 1)$ penalties. They also considered pairwise weights based on distances of observations in pairwise penalties.  The optimization problem was solved by the alternating direction method of multipliers algorithm (ADMM, \citet{boyd2011distributed}).  Using the ADMM algorithm,  the original optimization problem can be divided into several simpler sub-problems, which would be easier to solve.  This idea is extended to different regression settings. \cite{ma2017concave} and \cite{ma2016exploration} considered clustering problems in linear regression models using smoothly clipped absolute deviation (SCAD) penalty \citep{fan2001variable} and the minimax concave penalty (MCP) \citep{zhang2010nearly}. They also used the ADMM algorithm to solve the optimization problem constructed under linear regression models to find estimates of regression coefficients and the corresponding group structure. But they didn't consider pairwise weights in the penalty functions to incorporate extra information, such as locations. However, some extra information can help find clusters that would be more reasonable and easier to interpret. For example, spatial location information is used to help find spatial continuous groups in \cite{wang2019spatial}. Age distances are used to find continuous age obesity groups in  \cite{miljkovic2021identifying}.  Both methods used weighted pairwise penalties  and the ADMM algorithm,  which can easily incorporate extra information in pairwise penalties. \cite{zhu2018cluster} used spline bases to represent mean functions and used the ADMM  algorithm to identify clusters for longitudinal data without estimating the covariance structure. \cite{fang2022biclustering} used the ADMM algorithm and spline functions for both clustering in both the sample and covariate dimensions. But they didn't consider any extra information, either. \cite{ma2022subgroup} considered the clustering problem for functional partial linear regression, not the mean functions.

In this work, a new method is proposed to use spatial information or ordering information to cluster longitudinal curves. The new approach can find cluster structures, estimate mean functions and the covariance function simultaneously using FPCA. In the proposed method, each individual curve is assumed to have its own mean function represented by B-spline bases \citep{de2001practical}. Eigenfunctions are also expressed by B-spline bases with some constraints in parameters,  which are assumed to be the same for all individual curves. Clusters of individual curves are identified based on the weighted pairwise concave penalty as in \cite{wang2019spatial}. In the proposed algorithm, spatial or location information can be considered when constructing pairwise weights if the information is known. Mean functions and the covariance function, along with the group structure, are estimated simultaneously by combining the ADMM algorithm and the EM algorithm.  The idea of the combination of the EM and the ADMM algorithm is also used in \cite{ren2021gaussian} and \cite{foulds2015latent} in other setups,  where the ADMM algorithm is used in the M-step in the EM algorithm.  
 \cite{zhou2021subgroup} proposed a two-stage algorithm, which used the ADMM algorithm in the first step to finding the initial values for the EM algorithm. And the difference between the proposed algorithm and the two-stage algorithm in \cite{zhou2021subgroup} is that  the ADMM algorithm is iteratively used in the EM algorithm instead of using it as an initial step. 

The contributions of this work can be summarized as below.  First,  a model based on FPCA with individual mean functions and weighted pairwise penalty functions  (FWP) is assumed,  which can incorporate extra spatial or location information.   Second,  a new algorithm is developed based on the EM and ADMM algorithms to find estimates and clusters. In both the simulation study and the application, data sets with regular time observations are considered.  The proposed method is compared to some existing methods in the simulation study.  The results show that the weighted penalty performs better if there is a potential spatial structure.

The article is organized as follows. In Section \ref{sec_model}, the FPCA model with individual mean functions and weighted pairwise penalty (FWP) is described. In Section \ref{sec_algorithm}, the proposed optimization problem and the algorithm are introduced. The simulation study is conducted in Section \ref{sec_simulation} to show the performance of the proposed method.  A real example is analyzed in  Section \ref{sec_realdata} to illustrate the new method.  Finally, some discussions are given in Section \ref{sec_summary}.

\section{The FPCA subgroup model}
\label{sec_model}
Following the model discussed in \cite{yao2005functional} and \cite{james2000principal}, let $\mathcal{T}$ be the time interval with $[0,1]$, and $Y_i(t)$ be the independent curves for $t\in \mathcal{T}$ and $i=1,\dots,n$. $X_{i}\left(t\right)$ is the latent functional process of $Y_i(t)$ and the covariance function is  {\rc $\Gamma_i\left(t_{1},t_{2}\right) =\text{Cov} [X_i(t_1),X_i(t_2)]$}.  Assume that $X_i(t)$ is a square integrable stochastic process over $\mathcal{T}$ with mean function $\mu_i(t)$ and  covariance function is continuous, then the covariance function can be decomposed as
$
\Gamma\left(t_{1},t_{2}\right)=\sum_{l=1}^{\infty}\lambda_{l}\psi_{l}\left(t_{1}\right)\psi_{l}\left(t_{2}\right),
$
where $\lambda_{1}\geq\lambda_{2}\geq\cdots>0$ are eigenvalues and $\psi_{l}\left(\cdot\right)$'s are corresponding eigenfunctions which are orthonormal, that is, $\int_{\mathcal{T}}\psi_{l}\left(t\right)\psi_{l^{\prime}}\left(t\right)dt=I\left(l=l^{\prime}\right)$. The covariance function here does not have the stationary assumption, and is more flexible.  Based on the Karhunen-Lo{\`e}ve expansion, $X_{i}\left(t\right)$ can be written as in (\ref{eq:kl_expansion}) 
\begin{equation}
\label{eq:kl_expansion}
X_{i}\left(t\right)=\mu_{i}\left(t\right)+\sum_{l=1}^{\infty}\xi_{il}\psi_{l}\left(t\right),
\end{equation}
where $\mu_i(t)$ is the mean function of the $i$th individual, $\xi_{il}$ is a normal random variable with {\rc $E\left[\xi_{il}\right]=0$ and $Var\left[\xi_{il}\right]=\lambda_{l}$}. Note that different individual curves have the same covariance function. In practice, it is not feasible to estimate the infinite number of components in the covariance function. Thus, the truncated form is used to approximate \eqref{eq:kl_expansion} as in \cite{james2000principal}, that is,
\begin{equation}
X_{i}\left(t\right) \approx   \mu_{i}\left(t\right)+\sum_{l=1}^{P}\xi_{il}\psi_{l}\left(t\right),
\label{eq:model_approx}
\end{equation}
where $P$ is the number of components that will be selected later. The method of choosing $P$ will be introduced in Section \ref{sec_simulation}. Then the model for $Y_i(t)$ is
\begin{equation}
Y_i(t) = \mu_{i}\left(t\right)+\sum_{l=1}^{P}\xi_{il}\psi_{l}\left(t\right)+\epsilon_i(t),
\label{eq:model_y}
\end{equation}
where $\epsilon_i(t)$ is the additional measurement error  following a normal distribution, which has mean $0$ and variance $\sigma^{2}$ and is independent of $\xi_{il}$. The Karhunen-Lo{\`e}ve expansion is also used in some other clustering work, such as \cite{jacques2013funclust} and \cite{huang2014joint}. In the previous work, latent variables were used to indicate the group information, which differs from the proposed method.  In the proposed method, the group information is indicated by values of parameters instead of latent variables, which will be introduced later in detail. 

Assume that both mean functions $\mu_i(t)$ and eigenfunctions $\psi_l(t)$ are smooth functions. Regression splines are used to approximate them. Specifically, let $\bm{B}\left(t\right)=\left(B_{1}\left(t\right),\dots,B_{q}\left(t\right)\right)^{T}$ be the $q$ dimensional B-spline bases with equally spaced knots defined on $\mathcal{T}$. Then, mean functions and eigenfunctions are expressed as
\begin{equation}
\label{eq_mean}
\mu_{i}\left(t\right)  =  \bm{B}^{T}\left(t\right)\bm{\beta}_{i},
\end{equation}
\begin{equation}
\label{eq_cov}
{\rc \left(\psi_{1}\left(t\right),\dots,\psi_{P}\left(t\right)\right)}  =  \bm{B}^{T}\left(t\right)\bm{\Theta},
\end{equation}
where $\bm{\beta}_i$'s are unknown coefficients, $\bm{\Theta}$ is a $q\times P$ parameter matrix.  Note that, $\bm{\beta}_i$'s indicate the difference among different mean functions. Let $\bm{\xi}_i = (\xi_{i1},\dots,\xi_{iP})^T$, then the reduced rank model becomes
\begin{equation}
\label{eq_bspline_model}
Y_i(t)=\bm{B}^{T}\left(t\right)\bm{\beta}_{i}+\bm{B}^{T}\left(t\right)\bm{\Theta}\bm{\xi}_{i}+\epsilon_i(t),
\end{equation}
where $\bm{\xi}_{i}\overset{iid}\sim N\left(\bm{0},\bm{\Lambda}\right)$ and $\bm{\Lambda}$ is a $P\times P$ diagonal matrix with the $l$th element as $\lambda_{l}$. As used in \cite{james2000principal} and \cite{zhou2008joint}, the orthogonality constraint of eigenfunctions $\psi_l(\cdot)$'s is guaranteed based on the following constraints on spline bases and the parameter matrix $\bm{\Theta}$, 
\begin{equation}
\int\bm{B}\left(t\right)\bm{B}\left(t\right)^{T}dt=\bm{I}_{q},\quad \bm{\Theta}^{T}\bm{\Theta}=\bm{I}_{P},
\label{eq:constraint_B}
\end{equation}
where $\bm{I}_{q}$ and $\bm{I}_{P}$ are a $q$-dimensional identity matrix and $P$-dimensional identity matrix, respectively. Instead of using the  numeric approximation procedure in \cite{zhou2008joint}, the matrix representation method is used to obtain the orthogonal B-spline bases functions \citep{redd2012comment}.   Based on Lemma 1 in \cite{zhou2008joint}, the identifiability of parameters $\bm{\Theta}$, $\bm{\Lambda}$ is guaranteed by two conditions: 1) $\bm{\Theta}^{T}\bm{\Theta}=\bm{I}_{P}$ and 2) the sign of the first element with the largest magnitude is positive in each column of $\bm{\Theta}$.

Let $t_{ih}$ for $h=1,\dots H$ be the observed time point,  and  $Y_i(t_{ih})$ be the observed value of $Y_i(t)$ at time $t_{ih}$.  Define $\bm{Y}_i = (Y_i(t_{i1}),\dots,Y_i(t_{iH}))^T$, $\bm{B}_i = \left(\bm{B}(t_{i1}),\dots, \bm{B}(t_{iH})\right)^T$ and $\bm{\epsilon}_i = (\epsilon_i(t_{i1}),\dots, \epsilon_i(t_{iH}))^T$, the data model becomes 
\begin{equation}
\label{eq:model}
\bm{Y}_i = \bm{B}_i\bm{\beta}_i+ \bm{B}_i\bm{\Theta}\bm{\xi}_i+\bm{\epsilon}_i.
\end{equation}
Assume that there are $K$ distinct groups with different mean functions, denoted by $\mathcal{G}_{1},\mathcal{G}_{2},\dots,\mathcal{G}_{K}$, which is a partition of $\{1, 2, \dots, n\}$. Under this partition, $\mu_{i}\left(t\right)= \mu_{j}\left(t\right)$ if $i$  and $j \in\mathcal{G}_{k}$, which means $i$ and $j$ are in the same group.  Based on the expression of the reduced rank model in \eqref{eq_bspline_model},  the clustering problem becomes to find a partition of $\{1,2,\dots, n \}$ such that $\bm{\beta}_i = \bm{\beta}_j$  if $i$ and $j$ are in the same group. But neither the partition $\mathcal{G}_k, k=1,2,\dots, K$ nor the number of clusters $K$ is known. Thus, the goal is to find the partition $\hat{\mathcal{G}}_k$ and the number of clusters $\hat{K}$ based on the observations.

To achieve the goal of estimating parameters and finding group structure, pairwise penalties are applied to the differences of $\bm{\beta}_i$ \citep{ma2016exploration,wang2019spatial}. Then, the following optimization problem is considered: minimize the objective function with pairwise penalties,
\begin{align}
\label{eq_obj00}
Q(\bm{\beta},\bm{\Theta},\bm{\xi},\bm{\lambda},\sigma^{2}) & =\frac{\tilde{\sigma}^2}{H}l\left(\bm{\beta},\bm{\Theta},\bm{\xi},\bm{\lambda},\sigma^{2}\right)+\sum_{1\leq i<j\leq n}p_{\gamma}\left(\Vert\bm{\beta}_{i}-\bm{\beta}_{j}\Vert,c_{ij}\tau\right),
\end{align}
where $\Vert \cdot \Vert$ denotes the Euclidean norm,  {\rc $\tilde{\sigma}^2$ is an initial value of $\sigma^2$}, and $l\left(\bm{\beta},\bm{\Theta},\bm{\xi},\bm{\lambda},\sigma^{2}\right)$ is the joint negative loglikelihood function similar to that in \cite{james2000principal}. The difference is that the proposed model has individual coefficients $\bm{\beta}_i$ instead of a common vector as that in \cite{james2000principal}.  $l\left(\bm{\beta},\bm{\Theta},\bm{\xi},\bm{\lambda},\sigma^{2}\right)$ has the following form, 
{\rc 
\begin{align}
\label{eq_lik}
l(\bm{\beta},\bm{\Theta},\bm{\xi},\bm{\lambda},\sigma^{2})= & \frac{1}{2\sigma^{2}}\sum_{i=1}^{n}\Vert \bm{Y}_{i}-\bm{B}_{i}\bm{\beta}_{i}-\bm{B}_{i}\bm{\Theta}\bm{\xi}_{i}\Vert^{2} \nonumber\\
+ & \frac{n}{2}\log\left(\vert\bm{\Lambda}\vert\right)+\frac{1}{2}\sum_{i=1}^{n}\bm{\xi}_{i}^{T}\bm{\Lambda}^{-1}\bm{\xi}_{i} + \frac{1}{2}\sum_{i=1}^{n}H\log(\sigma^{2}),
\end{align}
}
where $\bm{\beta} = (\bm{\beta}_1^T, \dots, \bm{\beta}_n^T)^T, \bm{\xi} = (\bm{\xi}_1^T,\dots, \bm{\xi}_n^T)^T, \bm{\lambda} = (\lambda_1,\dots, \lambda_P)^T$.  

The reason for using $\tilde{\sigma}^2$ is to avoid some numerical issues due to possible smaller values of $\sigma^2$ found in some simulations when having the second term of \eqref{eq_obj00} in the ADMM algorithm.   Adding a constant will not change the shape of the likelihood function. And in the algorithm, the parameter $\sigma^2$ will be estimated together with other parameters.  The method of obtaining $\tilde{\sigma}^2$ is discussed in Remark \ref{remark_initial} {\rc and the Appendix. \eqref{eq:sig20} in the Appendix gives the approach of obtaining $\tilde{\sigma}^2$.} The proposed method works well in different simulation setups.

In  \eqref{eq_obj00}, $p_{\gamma}\left(\cdot,\tau\right)$ is a penalty function with a tuning parameter $\tau \geq 0$, which will be selected later described in Section \ref{sec_simulation}. The penalty function can be $L_1$ penalty \citep{tibshirani1996regression}, SCAD \citep{fan2001variable} and MCP \citep{zhang2010nearly}.  In \cite{ma2017concave}, they considered both SCAD and MCP, and they showed that $L_1$ penalty tended to produce too many groups, SCAD and MCP performed similarly and had the same theoretical results. Thus, here we only consider the SCAD penalty, which is defined as,
\begin{equation}
\label{eq_scad}
p_{\gamma}(t,\lambda) = \lambda \int_0^{\vert t \vert} \min \{ 1,(\gamma - x/\lambda)_+/(\gamma - 1) \}dx.
\end{equation}
Here we treat $\gamma$ as a fixed value 3 as in \cite{ma2016exploration}.  As the value of $c_{ij}\tau$ increases, some pairs of $\Vert \bm{\beta}_i - \bm{\beta}_j\Vert$ would be shrunk to zeros, then $i$ and $j$ will be in the same group. According to the estimate of $\bm{\beta}$ denoted as $\hat{\bm{\beta}}$, we will have the estimated partition of $\{1,2,\dots, n\}$ such that $\hat{\bm{\beta}}_i = \hat{\bm{\beta}}_j$ if $i$ and $j$ are in the same group.

Besides that, an associated weight $c_{ij}$ is assigned to each pair of the penalty.  $c_{ij}$ is defined based on similarities between individual $i$ and individual $j$, such as distance, which is discussed in \cite{wang2019spatial}.  For closer locations, larger weights are assigned such that they tend to be grouped together. And for locations which are far away from each other, smaller weights are assigned and they tend to be separated.  For example,  if $\bm{s}_i$ is the location of individual $i$, then $c_{ij}$ can be defined as $c_{ij} = \exp(-\alpha\Vert \bm{s}_i - \bm{s}_j\Vert)$, where $\alpha$ is another tuning parameter to be chosen.  And it can be seen that when the distance between two locations is small, the corresponding weight $c_{ij}$ is large. With a larger weight,  these two locations tend to be shrunk together. 
Pairwise weights $c_{ij}$'s can be defined according to different contexts of different problems. In \cite{wang2019spatial}, they discussed several possible ways of defining the weights. The $c_{ij}$ is defined such that the largest value is 1 when $\alpha = 0$, which will not have a scale issue between $c_{ij}$ and $\tau$. In the simulation study, a spatial grid structure is considered, and corresponding weights are defined in Section \ref{subsec_secnario2}. Both $\tau$ and $\alpha$ are tuning parameters, which will be selected based on Bayesian information criteria (BIC) \citep{ma2017concave,wang2019spatial}. The detail will be given in Section \ref{sec_simulation}.

\section{The proposed  algorithm}
\label{sec_algorithm}

In this section, the proposed algorithm is introduced in detail. The proposed algorithm is an iterative algorithm, combining the EM and ADMM algorithms.  

Recall that the goal is to minimize the objective function proposed in \eqref{eq_obj00}. \eqref{eq_obj0} shows more details about the objective function,
\begin{align}
\label{eq_obj0}
Q(\bm{\beta},\bm{\Theta},\bm{\xi},\bm{\lambda},\sigma^{2}) &  =\frac{\tilde{\sigma}^2}{2H\sigma^{2}}\sum_{i=1}^{n}\Vert\bm{Y}_{i}-\bm{B}_{i}\bm{\beta}_{i}-\bm{B}_{i}\bm{\Theta}\bm{\xi}_{i}\Vert^{2} \nonumber\\
 & +\frac{\tilde{\sigma}^2}{2}\sum_{i=1}^{n}\log(\sigma^{2})+\frac{n\tilde{\sigma}^2}{2H}\log\left(\vert\bm{\Lambda}\vert\right)+\frac{\tilde{\sigma}^2}{2H}\sum_{i=1}^{n}\bm{\xi}_{i}^{T}\bm{\Lambda}^{-1}\bm{\xi}_{i} \nonumber\\
 & +\sum_{1\leq i<j\leq n}p_{\gamma}\left(\Vert\bm{\beta}_{i}-\bm{\beta}_{j}\Vert,c_{ij}\tau\right).
\end{align}

The ADMM algorithm is used widely to solve this type of optimization problem in linear regression setups with pairwise penalties. In the ADMM algorithm,  let $\bm{\delta}_{ij}=\bm{\beta}_{i}-\bm{\beta}_{j}$, then the objective function becomes
\begin{align*}
Q_{0}(\bm{\beta},\bm{\Theta},\bm{\xi},\bm{\lambda},\sigma^{2},\bm{\delta}) & =\frac{\tilde{\sigma}^2}{2H\sigma^{2}}\sum_{i=1}^{n}\Vert\bm{Y}_{i}-\bm{B}_{i}\bm{\beta}_{i}-\bm{B}_{i}\bm{\Theta}\bm{\xi}_{i}\Vert^{2}\\
 & +\frac{\tilde{\sigma}^2}{2}\sum_{i=1}^{n}\log(\sigma^{2})+\frac{n\tilde{\sigma}^2}{2H}\log\left(\vert\bm{\Lambda}\vert\right)+\frac{\tilde{\sigma}^2}{2H}\sum_{i=1}^{n}\bm{\xi}_{i}^{T}\bm{\Lambda}^{-1}\bm{\xi}_{i}\\
 & +\sum_{1\leq i<j\leq n}p_{\gamma}\left(\Vert\bm{\delta}_{ij}\Vert,c_{ij}\tau\right)\\
 & \text{subject to }\bm{\beta}_{i}-\bm{\beta}_{j}-\bm{\delta}_{ij}=0.
\end{align*}
The augmented Lagrangian without the constraints is 
\begin{align}
\label{eq:target-1}
Q_{1}\left(\bm{\beta},\bm{\Theta},\bm{\xi},\bm{\lambda},\sigma^{2},\bm{\delta},\bm{v}\right) & =Q_{0}\left(\bm{\beta},\bm{\Theta},\bm{\xi},\bm{\lambda},\sigma^{2},\bm{\delta}\right)+\sum_{i<j}\left\langle \bm{v}_{ij},\bm{\beta}_{i}-\bm{\beta}_{j}-\bm{\delta}_{ij}\right\rangle \nonumber \\
 & +\frac{\vartheta}{2}\sum_{i<j}\Vert\bm{\beta}_{i}-\bm{\beta}_{j}-\bm{\delta}_{ij}\Vert^{2},
\end{align}
where $\bm{v}=\left(\bm{v}_{ij}^{T}, i<j\right)^{T}$ are Lagrange multipliers and $\vartheta$ is the penalty parameter, which is fixed at 1 here as in \cite{ma2017concave} and \cite{ma2016exploration}.  Parameters will be updated based on minimizing the objective function \eqref{eq:target-1}.

As in \cite{james2000principal} and \cite{zhou2008joint},  the EM algorithm can be used when treating $\bm{\xi}_i$'s as missing data. In the E-step of the EM algorithm, the distribution of  $\bm{\xi}_i$ is needed given the current values of parameters.  From \cite{james2000principal},  the conditional distribution of $\bm{\xi}_i$ is
\[
\bm{\xi}_{i}\vert \bm{\beta}, \bm{\Theta},\bm{\lambda},\sigma^2 \sim N\left(\bm{m}_i, \bm{V}_i\right),
\]
where
\begin{equation}
\label{eq:condmean}
\bm{m}_{i}= {\rc E\left[\bm{\xi}_{i}\vert  \bm{\beta}, \bm{\Theta},\bm{\lambda},\sigma^2  \right]}=\left(\bm{\Theta}^{T}\bm{B}_{i}^{T}\bm{B}_{i}\bm{\Theta}+\sigma^{2}\bm{\Lambda}^{-1}\right)^{-1}\bm{\Theta}^{T}\bm{B}_{i}^{T}\left(\bm{Y}_{i}-\bm{B}_{i}\bm{\beta}_{i}\right),
\end{equation}
\begin{equation}
\label{eq:condvar}
\bm{V}_{i}= {\rc V\left[\bm{\xi}_{i}\vert  \bm{\beta}, \bm{\Theta},\bm{\lambda},\sigma^2  \right]}=\left(\frac{1}{\sigma^{2}}\bm{\Theta}^{T}\bm{B}_{i}^{T}\bm{B}_{i}\bm{\Theta}+\bm{\Lambda}^{-1}\right)^{-1}.
\end{equation}
Note that, in \cite{james2000principal},  a common mean function is used. Here individual $\bm{\beta}_i$s are used for individual mean functions.  When all individuals have the same observed time points, then all $\bm{V}_i$'s are the same here. This is the case considered in this article. For simplicity, $\bm{V}$ will be used to denote $\bm{V}_i$ later.  Define $\hat{\bm{m}}_{i}$ and $\hat{\bm{V}}$ as the  evaluated vector and matrix of $\bm{m}_i$ and $\bm{V}$ at $\hat{\bm{\beta}},\hat{\bm{\Theta}},\hat{\bm{\lambda}}$ and $\hat{\sigma}^2$, respectively.  
The proposed algorithm can be extended to a more general case. 
 
In the M-step,  parameters $\bm{\beta},\bm{\Theta}, \bm{\lambda}, \sigma^2, \bm{\delta}$ and Lagrange multipliers $\bm{v}$ will be updated.   The details will be introduced in two parts.  In part 1, $\bm{\Theta}$, $\bm{\lambda}$ and $\sigma^2$ are updated,  and all of these updates don't depend on the penalty part.  In part 2,  $\bm{\beta}$, $\bm{\delta}$ and $\bm{v}$ are updated. 
\subsection*{Part 1} 
Let $\bm{\beta}^{(r)}$, $\bm{\Theta}^{(r)}$, $\bm{\lambda}^{(r)}$,  $(\sigma^2)^{(r)}$,  $\bm{\delta}^{(r)}$ and $\bm{v}^{(r)}$ be estimates at the $r$-th iteration. 
$\sigma^2$ is updated based on $E\left[Q_{1}\left(\bm{\beta},\bm{\Theta},\bm{\xi},\bm{\lambda},\sigma^{2},\bm{\delta},\bm{v}\vert\bm{\beta}^{(r)},  \bm{\Theta}^{(r)},\bm{\lambda}^{(r)},(\sigma^2)^{(r)} ,  \bm{\delta}^{(r)}, \bm{v}^{(r)}\right)\right]$, that is 
\begin{align}
\label{eq:updatesig2}
\left(\sigma^{2}\right)^{(r+1)} & =\frac{1}{nH}\sum_{i=1}^{n}\Vert\bm{Y}_{i}-\bm{B}_{i}\bm{\beta}_{i}^{(r)}-\bm{B}_{i}\bm{\Theta}^{(r)}\hat{\bm{m}}_{i}\Vert^{2} \nonumber\\
 & +\frac{1}{nH}\sum_{i=1}^{n}tr\left(\bm{B}_{i}\bm{\Theta}^{(r)}\hat{\bm{V}}\left(\bm{\Theta}^{(r)}\right)^{T}\bm{B}_{i}^{T}\right).
\end{align}

When updating $\bm{\Theta}$, each column is updated sequentially as in \cite{zhou2008joint} and \cite{huang2014joint}.  Let $\bm{\theta}_j$ be the $j$th column of $\bm{\Theta}$ for $j=1,\dots , P$. $\bm{\theta}_j$ is updated by minimizing the following expectation with respect to $\bm{\theta}_{j}$,
\[
E\left[\sum_{i=1}^{n}\left(\bm{Y}_{i}-\bm{B}_{i}\bm{\beta}_{i}^{(r)}-\sum_{l\neq j}\bm{B}_{i}\bm{\theta}_{l}\xi_{il}-\bm{B}_{i}\bm{\theta}_{j}\xi_{ij}\right)^{2}\Big\vert\bm{\beta}^{(r)}, \bm{\Theta}^{(r)}, \bm{\lambda}^{(r)}, (\sigma^2)^{(r)}\right].
\]
Therefore, the estimate for $\bm{\theta}_j$ is 
\begin{align}
\label{eq:updatetheta}
\tilde{\bm{\theta}}_{j} & =\left(\sum_{i=1}^{n}\bm{B}_{i}^{T}\bm{B}_{i}\left(\hat{m}_{ij}^{2}+\hat{\bm{V}}\left(j,j\right)\right)\right)^{-1}\nonumber \\
 & \cdot\sum_{i=1}^{n}\bm{B}_{i}^{T}\left[\left(\bm{Y}_{i}-\bm{B}_{i}\bm{\beta}_{i}^{(r)}\right)\hat{m}_{ij}-\sum_{l\neq j}\bm{B}_{i}\bm{\theta}_{l}\left(\hat{m}_{il}\hat{m}_{ij}+\hat{\bm{V}}\left(l,j\right)\right)\right],
\end{align}
where $\hat{m}_{ij}$ is the $j$th element of $\hat{\bm{m}}_{i}$ and $\hat{\bm{V}}(l,j)$ is the $lj$th element of $\hat{\bm{V}}$. But the matrix $\tilde{\bm{\Theta}} = \left(\tilde{\bm{\theta}}_1, \dots, \tilde{\bm{\theta}}_P\right)$ obtained by this procedure is not orthonormal.  The same procedure  as in \cite{zhou2008joint}  and  \cite{huang2014joint} is used to orthogonalize $\bm{\Theta}$ and provide the updated estimate of $\bm{\lambda}$. In this procedure, compute
$
\bm{\Sigma}=\frac{1}{n}\sum_{i=1}^{n}\left(\hat{\bm{m}}_{i}\hat{\bm{m}}_{i}^{T}+\hat{\bm{V}}\right).
$
Then an eigenvalue decomposition is done such that $\tilde{\bm{\Theta}}\bm{\Sigma}\tilde{\bm{\Theta}}^{T}=\hat{\bm{\Theta}}\hat{\bm{\Lambda}}\hat{\bm{\Theta}}^{T}$, where $\hat{\bm{\Lambda}}$ is a diagonal matrix with eigenvalues
arranged in decreasing order and $\hat{\bm{\Theta}}$ has the corresponding
eigenvectors. Thus, $\hat{\bm{\Lambda}}$ is the update of $\bm{\Lambda}$ denoted as $\bm{\Lambda}^{(r+1)}$
and $\hat{\bm{\Theta}}$ is the update of $\bm{\Theta}$ denoted as $\bm{\Theta}^{(r+1)}$.

By applying the above procedure,  updates of parameters,  $\bm{\Theta}^{(r+1)}$, $\bm{\Lambda}^{(r+1)}$ and $(\sigma^2)^{(r+1)}$ are obtained.  

\subsection*{Part 2}
 When ignoring other unrelated components about  $\bm{\beta}$, $\bm{\delta}$ and $\bm{v}$ in \\$E\left[Q_{1}\left(\bm{\beta},\bm{\Theta},\bm{\xi},\bm{\lambda},\sigma^{2},\bm{\delta},\bm{v}\vert\bm{\beta}^{(r)},  \bm{\Theta}^{(r)},\bm{\lambda}^{(r)},(\sigma^2)^{(r)} ,  \bm{\delta}^{(r)}, \bm{v}^{(r)}\right)\right]$,  the objective function becomes, 
\begin{align}
\label{eq:target-2}
 Q_{2}\left(\bm{\beta},\bm{\delta},\bm{v}\right) & = \frac{\tilde{\sigma}^2}{2H\left(\sigma^{2}\right)^{(r+1)}}\sum_{i=1}^{n}\Vert \bm{Y}_{i}-\bm{B}_{i}\bm{\beta}_{i}-\bm{B}_{i}\bm{\Theta}^{(r+1)}\hat{\bm{m}}_{i}\Vert^{2} \nonumber\\
 & +\sum_{i<j}\left\langle \bm{v}_{ij},\bm{\beta}_{i}-\bm{\beta}_{j}-\bm{\delta}_{ij}\right\rangle +\frac{\vartheta}{2}\sum_{i<j}\Vert\bm{\beta}_{i}-\bm{\beta}_{j}-\bm{\delta}_{ij}\Vert^{2} \nonumber \\
 & + \sum_{i<j}p_{\gamma}\left(\Vert\bm{\delta}_{ij}\Vert,c_{ij}\tau\right).
\end{align}

Based on \eqref{eq:target-2}, $\bm{\beta}$, $\bm{\delta}$ and $\bm{v}$ are updated as follows to minimize the above objective function.  $\bm{\beta}$ is updated as below,
{\rc
\begin{align}
\label{eq:updatebeta}
\bm{\beta}^{(r+1)} & =\left(\bm{B}_{0}^{T}\bm{B}_{0}+\vartheta \left(\tilde{\sigma}^2\right)^{-1}H(\sigma^{2})^{(r+1)}\bm{A}^{T}\bm{A}\right)^{-1}\left[\bm{B}_{0}^{T}\left(\bm{Y}-\tilde{\bm{B}}\hat{\bm{m}}\right)\right.\nonumber\\
 & +\left.\vartheta \left(\tilde{\sigma}^2\right)^{-1} H(\sigma^{2})^{(r+1)}\text{vec}\left(\left(\bm{\Delta}^{(r)}-\vartheta^{-1}\bm{\Upsilon}^{(r)}\right)\bm{D}\right)\right]
\end{align}
}
where $\bm{B}_{0}=\text{diag}\left(\bm{B}_{1},\dots,\bm{B}_{n}\right)$,
$\tilde{\bm{B}}=\text{diag}\left(\bm{B}_{1}\bm{\Theta}^{(r+1)},\dots,\bm{B}_{n}\bm{\Theta}^{(r+1)}\right)$,
$\bm{Y}=\left(\bm{Y}_{1}^{T},\dots,\bm{Y}_{n}^{T}\right)^{T}$, $\hat{\bm{m}}=\left(\hat{\bm{m}}_{1}^{T},\dots,\hat{\bm{m}}_{n}^{T}\right)^{T}$,
$\bm{A}=\bm{D}\otimes\bm{I}_{q}$ ($\otimes$ is the kronecker product), $\bm{D}=\left\{ \left(\bm{e}_{i}-\bm{e}_{j}, i<j\right)^{T}\right\} $, $\bm{e}_i$ is an $n \times 1$ vector with $i$th element 1 and other elements 0, {\rc $\bm{\Delta}=\left(\bm{\delta}_{ij}^{(r)},i<j\right)$ is a $q\times n\left(n-1\right)/2$ matrix and $\bm{\Upsilon}^{(r)}=\left(\bm{v}_{ij}^{(r)},i<j\right)$ is a $q\times n\left(n-1\right)/2$ matrix.}

$\bm{\delta}$  is updated  by minimizing 
\begin{equation*}
\frac{\vartheta}{2}\Vert\bm{\varsigma}_{ij}^{(r)}-\bm{\delta}_{ij}\Vert^{2}+p_{\gamma}\left(\Vert\bm{\delta}_{ij}\Vert,c_{ij}\tau\right),
\end{equation*}
 where $\bm{\varsigma}_{ij}^{(r)}=\left(\bm{\beta}_{i}^{(r+1)}-\bm{\beta}_{j}^{(r+1)}\right)+\vartheta^{-1}\bm{v}_{ij}^{(r)}$.
The solution based on the SCAD penalty is 
\begin{equation}
\label{eq:updatedelta}
\bm{\delta}_{ij}^{(r+1)}=\begin{cases}
S\left(\bm{\varsigma}_{ij}^{(r)},\tau c_{ij}/\vartheta\right) & \text{if }\left\Vert \bm{\varsigma}_{ij}^{(r)}\right\Vert \leq\tau c_{ij}+\tau c_{ij}/\vartheta,\\
\frac{S\left(\bm{\varsigma}_{ij}^{(r)},\gamma \tau c_{ij}/\left(\left(\gamma-1\right)\vartheta\right)\right)}{1-1/\left(\left(\gamma-1\right)\vartheta\right)} & \text{if }\tau c_{ij}+\tau c_{ij}/\vartheta<\left\Vert \bm{\varsigma}_{ij}^{(r)}\right\Vert \leq\gamma\tau c_{ij},\\
\bm{\varsigma}_{ij}^{(r)} & \text{if }\left\Vert \bm{\varsigma}_{ij}^{(r)}\right\Vert >\gamma\tau c_{ij},
\end{cases}
\end{equation}
where $\gamma>1+1/\vartheta$, $S\left(\bm{w},t\right)=\left(1-t/\Vert\bm{w}\Vert\right)_{+}\bm{w}$
and $\left(t\right)_{+}=t$ if $t>0$, $0$ otherwise.  Finally, $\bm{v}_{ij}$ is updated as in the typical ADMM algorithm,
\begin{equation}
\label{eq:updatev}
\bm{v}_{ij}^{(r+1)} =\bm{v}_{ij}^{(r)}+\vartheta\left(\bm{\beta}_{i}^{(r+1)}-\bm{\beta}_{j}^{(r+1)}-\bm{\delta}_{ij}^{(r+1)}\right).
\end{equation}

The proposed algorithm can be summarized as follows. 

\begin{algorithm}[H]
	\caption*{{\bf{Algorithm}:} The EM-ADMM algorithm}
	\begin{algorithmic}[1]
	    \REQUIRE: Initialize $\bm{\beta}^{(0)}$, $\bm{\delta}^{(0)}$, $\bm{v}^{(0)}$, $\bm{\Theta}^{(0)}$, $(\sigma^2)^{(0)}$ and $\bm{\lambda}^{(0)}$. 
	    \FOR {$r=1,2,\dots$}
		\STATE Calculate $\hat{\bm{m}_i}$ and $\hat{\bm{V}}$ for $i=1,\dots, n$ according to \eqref{eq:condmean} and \eqref{eq:condvar}.
		\STATE Update $\sigma^2$ by \eqref{eq:updatesig2}.
		\STATE Calculate $\tilde{\bm{\Theta}}$ by \eqref{eq:updatetheta}.
		\STATE Update $\bm{\Theta}$ and $\bm{\lambda}$ by the orthonormal procedure. 
		\STATE Update $\bm{\beta}$ by \eqref{eq:updatebeta}.
		\STATE Update $\bm{\delta}$ by \eqref{eq:updatedelta} 
		\STATE Update $\bm{v}$ by \eqref{eq:updatev}.
		\IF{convergence criterion is met} 
		\STATE{Stop and get the estimates}
		\ELSE
		\STATE { $r=r+1$}
		\ENDIF
		\ENDFOR
	\end{algorithmic}
\end{algorithm}

\begin{remark}
The stopping criterion is based on the criterion in \cite{boyd2011distributed}. Define 
\[
\bm{u}^{(r+1)}=\bm{A}\bm{\beta}^{(r+1)}-\bm{\delta}^{(r+1)},
\]
and
\[
\bm{s}^{(r+1)}=\vartheta\bm{A}^{T}\left(\bm{\delta}^{(r+1)}-\bm{\delta}^{(r)}\right).
\]The stopping criterion is 
\[
\left\Vert \bm{u}^{(r)}\right\Vert \leq\epsilon^{\text{pri}},\quad\left\Vert \bm{s}^{(r)}\right\Vert \leq\epsilon^{\text{dual}}
\] 
with
\begin{align*}
\epsilon^{\text{pri}} & =\sqrt{\frac{n\left(n-1\right)}{2}q}\epsilon^{\text{abs}}+\epsilon^{\text{rel}}\max\left\{ \left\Vert \bm{A}\bm{\beta}^{(r)}\right\Vert,\left\Vert \bm{\delta}^{(r)}\right\Vert \right\}, \\
\epsilon^{\text{dual}} & =\sqrt{nq}\epsilon^{\text{abs}}+\epsilon^{\text{rel}}\left\Vert \bm{A}^{T}\bm{v}^{(r)}\right\Vert.
\end{align*}
The values of $\epsilon^{\text{abs}}$ and $\epsilon^{\text{rel}}$ are $10^{-4}$ and $10^{-2}$, respectively.
\end{remark}

\begin{remark}
\label{remark_initial}
The following procedure is used to initialize starting values of parameters. First calculate the  coefficients for each individual using the following form, 
\[
\text{\ensuremath{\bm{\beta}_{i}^{*}=\bm{B}_{i}\left(\bm{B}_{i}^{T}\bm{B}_{i}+\tau_{1}\Omega_{1}\right)^{-1}\bm{B}_{i}^{T}\bm{Y}_{i}}},
\]
where $\Omega_1 = \int \bm{B}(t) \bm{B}(t)^Tdt$ is the roughness penalty, which is an identity matrix here due to the constraint of the basis function in \eqref{eq:constraint_B}. Generalized cross validation (GCV) is used to select $\tau_1$ with the following form
\[
GCV\left(\text{\ensuremath{\tau_{1}}}\right)=\sum_{i=1}^{n}\frac{H\bm{Y}_{i}^{T}\left(\bm{I}-\bm{L}_{i}\right)^{2}\bm{Y}_{i}}{tr\left[\bm{I}-\text{\ensuremath{\bm{L}_{i}}}\right]^{2}},
\]
where $\bm{L}_{i}=\bm{B}_{i}\left(\bm{B}_{i}^{T}\bm{B}_{i}+\tau_{1}\Omega_{1}\right)^{-1}\bm{B}_{i}^{T}$. 

Then, $k$-means is used to obtain an initial group information based on initial estimates $\bm{\beta}_{i}^{*}$, given the number of groups. According to the given group structure by $k$-means,  the EM algorithm is then applied to obtain  initial values of $\bm{\beta}$, $\bm{\Theta}$, $\bm{\lambda}$ and $\sigma^2$.  The EM algorithm is  presented in the Appendix. $\bm{\delta}^{(0)}$ is initialized as $\bm{\delta}^{(0)} = \bm{A}\bm{\beta}^{(0)}$ and $\bm{v}$ is initialized as $\bm{0}$ for each element.
\end{remark}

As shown above, the proposed algorithm can estimate the group structure and mean functions by estimating $\bm{\beta}$. And estimated values of $\bm{\Theta}$ and $\bm{\lambda}$ give the estimated covariance function. 

Here, a two-step procedure is used to select the number of components $P$, tuning parameters $\tau$ and $\alpha$. A two-step procedure is commonly used when there are multiple tuning parameters, examples can be found in \cite{zhu2018cluster} and \cite{zhang2022subgroup}. In the first step, $P$ is selected when fixing $\tau = 0$ and $\alpha = 0$. In particular, $P$ components are used such that at least  95\% variation is explained, this is method is widely used in functional data analysis, see examples in \cite{james2000principal,xiao2022partial,zhang2022subgroup}.
When $P$ is selected, $\tau$ and $\alpha$ will be selected based on the selected $P$. 
Let $\hat{K}(\tau,\alpha)$ be the number of estimated groups for given values of $\tau$ and $\alpha$.  The following modified Bayesian information criteria (BIC) is used to select $\tau$ and $\alpha$, which adopts the forms in \cite{li2013selecting} and \cite{wang2007tuning}. The modified BIC is defined as 
\begin{equation}
\label{eq:BIC}
BIC(\tau, \alpha) = 2\hat{l}+C_{n}\log\left(nH\right){\rc \left(\hat{K} (\tau,\alpha) q\right)},
\end{equation}
where $\hat{l}$ is $E[l]$ in \eqref{eq_lik} evaluated at the estimates, and  $2\hat{l}=nH\log\left(\hat{\sigma}_{P}^{2}\right)+n\sum_{l=1}^{P}\log\hat{\lambda}_{l}+\sum_{i=1}^{n}\hat{\bm{m}}_{i}^{T}\hat{\bm{\Lambda}}^{-1}\hat{\bm{m}}_{i}$, which is typically used in EM based algorithm \citep{ibrahim2008model,huang2014joint}. When $C_n=1$, it becomes a traditional BIC. But when pairwise penalty is used, a modified $C_n$ is usually used in different models. Here $C_n = \log(\log(n))$ is used as in \cite{ma2017concave,zhang2022subgroup,li2021clusterwise}. $\hat{\sigma}_P^2$ {\rc based on $P$ components} has the following form.
$$
{\rc \hat{\sigma}_P^2} = \frac{1}{nH}\left( \sum_{i=1}^n \Vert \bm{Y}_i  - \bm{B}_i\hat{\bm{\beta}}_i - \bm{B}_i\hat{\bm{\Theta}}\hat{\bm{m}}_i \Vert^2 \right).
$$
When $ c_{ij}\tau$ increases, more pairs of $\Vert\bm{\beta}_{i}-\bm{\beta}_{j}\Vert$ will become 0, thus, the group structure can be estimated,  together with the number of clusters. By selecting $\tau$ and $\alpha$, the number of clusters and the group structure will be selected. When selecting tuning parameters, a two-dimensional grid search was used as in \cite{wang2019spatial}, where a grid search method is widely used in penalty-based approach \citep{ma2017concave,ma2022subgroup,fang2022biclustering, tibshirani1996regression,fan2001variable}. In particular, a grid of values of $\tau$ and a grid of values of $\alpha$ are predefined. The combination of $\tau$ and $\alpha$ with the smallest BIC is selected, then the number of clusters and the cluster structure are determined. 
The proposed method works well in terms of selecting the number of clusters and the number of components in the simulation study.  
The number of knots used is based on $q \approx (nH )^{1/5} + 4$ in \cite{huang2014joint} and \cite{li2010uniform} or some values around this value. The algorithm can be found in \url{https://github.com/wangx23/FWP}.

\section{Simulation study}
\label{sec_simulation}

In this section, the simulation study is conducted to compare the performance of the proposed new method FWP to some existing methods.

For each individual curve $i$, the same time points are considered with $t_{ih} = h/( H+1)$ for $h = 1,2,\dots,  H$ without boundary points, where $H$ is the total number of observations for each individual. Data sets are simulated based on the below model with three or four groups and two principal components,
$$
Y_i(t_{ih}) = X_i(t_{ih}) + \epsilon_i(t_{ih})\quad \text{with } X_{i}\left(t\right)=\mu_{i}\left(t\right)+\sum_{l=1}^{2}\xi_{il}\psi_{l}\left(t\right).
$$
Several sets of mean functions are considered in the simulation study. Two principal components are considered with $\psi_1(t) = \sqrt{2}\sin \left( 2\pi t\right)$ and  $\psi_2(t) = \sqrt{2}\cos \left( 2\pi t\right)$.  And $\xi_{il} \overset{iid}{\sim} N(0, \lambda_l)$  for $l=1, 2$ with  $\lambda_1 = 0.1$ and $\lambda_2 = 0.2$, $\epsilon_i(t_{ih}) \overset{iid}{\sim} N(0,\sigma^2)$ with  $\sigma = 0.2$.

In the simulation study, $H=10, 20, 30$ are considered and the number of knots are 7, 9 and 9, respectively. 
To evaluate the performance of the proposed method, the estimated group number $\hat{K}$, adjusted Rand index (ARI) \citep{rand1971objective, hubert1985comparing, vinh2010information} are reported.  The ARI measures the degree of agreement between two partitions,  taking the largest value 1: the larger ARI value, the more agreement.  The performance of estimating the curve is defined as follows
\begin{equation}
\label{eq:rmse}
RMSE = \sqrt{\frac{1}{n}\sum_{i=1}^n\Vert \hat{\bm{\mu}}_i - \bm{\mu}_i \Vert^2},
\end{equation}
where $\hat{\bm{\mu}_i} = \bm{B}_i\hat{\bm{\beta}}_i$ and $\bm{\mu}_i$ is the true curve mean from the simulation setting. The average $\hat{K}$ and the average ARI over 100 simulations are reported along with values of standard deviation in the parenthesis.  When using the proposed method, the selected number of components is also reported with the average value along with the standard deviation.  

Several methods are compared to the proposed method.  ``IND" represents the model without covariance structure proposed in \cite{zhu2018cluster},  ``JS" represents the method proposed in \cite{james2003clustering}  and ``FWP" represents the proposed method. The number of clusters for ``JS" is determined based on the ``distortion function" approach described in \cite{james2003clustering} and \cite{sugar2003finding}.  Besides these,  three other methods in functional data clustering with available R packages are also included. ``FEM" represents the method proposed in \cite{bouveyron2015discriminative}, which is implemented by R package {\it funFEM}.  ``funHDDC" represents the method proposed in \cite{bouveyron2011model}, which is implemented by the R package {\it funHDDC }. ``KMA" represents the method proposed in \cite{sangalli2010k}, which is implemented by the R package {\it fdacluster}.  For ``FEM" and ``funHDDC", BIC methods provided in packages are used to select the number of clusters and other potential structures in models.  For ``KMA", the number of clusters is fixed at the true number of clusters. The number of knots used in these methods are the same as the proposed method. 


\subsection{Scenario 1}

In this scenario, a random group structure with three groups is considered. Each group has 50 individuals, and each individual has the same number of time points (denoted as $H$). Three values of $H$ are considered with $10, 20, 30$. 
The mean functions of these three groups are $\mu_1(t) = \sqrt{2}\sin(4\pi t)$, $\mu_2(t) =  \exp(-10(t-0.25)^2)$ and $\mu_3(t) = 1.5t - 1$. If individual $i$ is in group $k$ for $k=1,2,3$, then $\mu_i(t) = \mu_k(t)$.

\begin{figure}[h]
\centering
\includegraphics[scale=0.6]{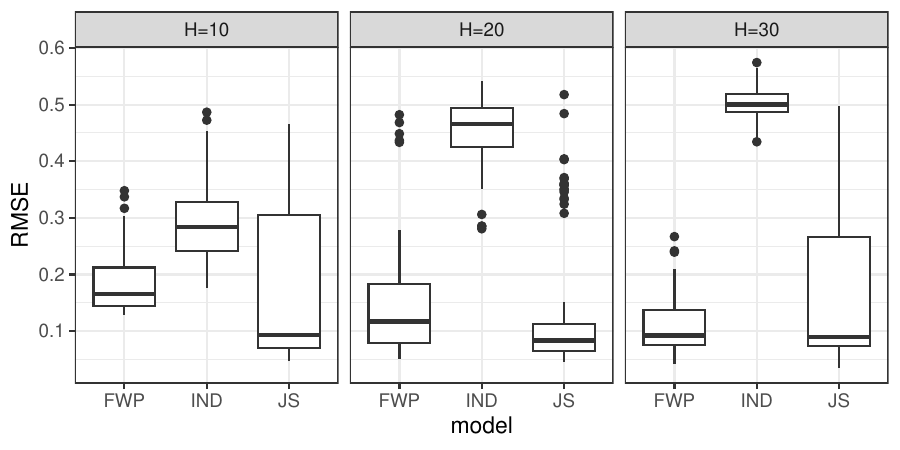}
\caption{RMSE of Scenario 1}
\label{fig:rmse_setting1}
\end{figure}

Table \ref{tab_KR_setting1} and Figure \ref{fig:rmse_setting1} show the results about the estimated number of groups $\hat{K}$, ARI and the number of components. From the results, ``FWP" performs better than ``IND"in terms of estimating the number of groups, recovering the group structure (large ARI) and estimating the mean functions (small RMSE).   ``FWP" is also better than  ``JS", ``FEM", ``funHDDC" and ``KMA" in terms of estimating the number of groups and recovering the true group structure. 
As $H$ increases the performance of ``FWP" becomes better, but not for ``IND".  Besides that, the number of components can be selected well.  In the simulation study of \cite{zhu2018cluster}, they showed that when the covariance structure is AR(1) or exchangeable, the method they proposed without covariance structure can capture the group structure well. However, under the setup here with a more flexible covariance structure, the performance becomes worse.

\begin{table}[h]
\centering
\caption{ Summary of $\hat{K}$ and average ARI for Scenario 1}
\label{tab_KR_setting1}
\begin{tabular}{cc|ccc}
\hline
& &  $H=10$ & $H=20$ & $H = 30$\\
\hline
 \multirow{5}{*}{$\hat{K}$} & IND & 7.78(2.81) & 20.11(5.81) & 28.62(5.76) \\ 
&  FWP & 3.44(0.76) & 3.84(2.02) & 3.25(0.48) \\ 
&  JS & 3.9(1.4) & 3.45(1.08) & 3.61(1.25) \\ 
&  FEM & 2.26(0.61) & 3.98(0.43) & 4.22(0.7) \\ 
&  HDDC & 4.15(0.87) & 3.81(0.95) & 4.34(0.87) \\ 
\hline 
\multirow{6}{*}{ARI} &  IND & 0.9(0.089) & 0.5(0.162) & 0.31(0.073) \\ 
 & FDA & 0.97(0.04) & 0.95(0.142) & 0.99(0.018) \\ 
 & JS & 0.89(0.168) & 0.94(0.14) & 0.9(0.181) \\ 
 & FEM & 0.55(0.038) & 0.44(0.097) & 0.64(0.216) \\ 
 & HDDC & 0.43(0.09) & 0.6(0.151) & 0.67(0.141) \\ 
 & KMA & 0.75(0.147) & 0.82(0.132) & 0.83(0.136) \\ 
\hline 
components & FWP &1.85(0.36) &2(0) &2(0)  \\
\hline
\end{tabular}
\end{table}

\subsection{Scenario 2}
\label{subsec_secnario2}

\begin{figure}[H]
\centering
\includegraphics[scale=0.4]{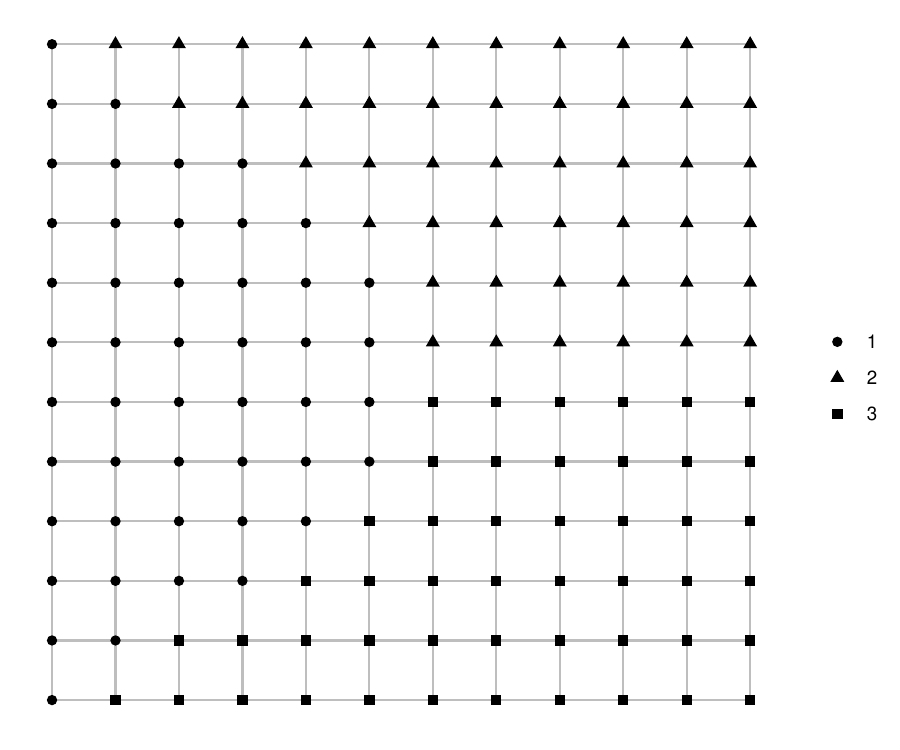}
\caption{Spatial group structure}
\label{fig:groupsp}
\end{figure}

The second scenario is considered in a grid lattice with a spatial group structure as shown in Figure \ref{fig:groupsp}. Each dot represents an individual with an associated location. If two locations are connected or neighbors, then the grey line is connected to these two dots. And the three different shapes represent three groups. There are 48 individuals in each group. The same mean functions are used as in Scenario 1. 

\begin{figure}[H]
\centering
\includegraphics[scale=0.25]{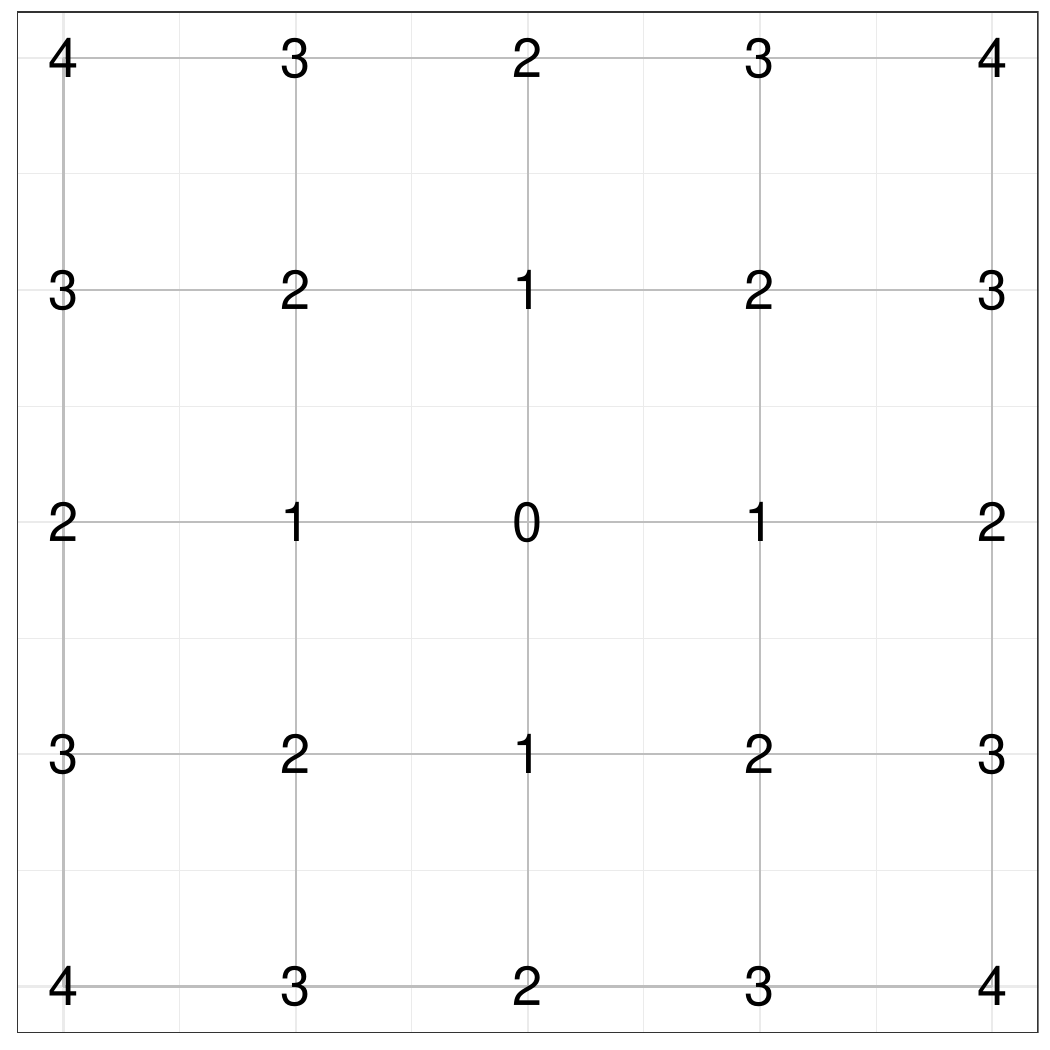}
\caption{An example of the definition of neighbor order}
\label{fig:neighboor}
\end{figure}

As mentioned in Section \ref{sec_algorithm}, pairwise weights are considered in this Scenario. The pairwise weights have the following form as used in \cite{wang2019spatial},
\begin{equation}
\label{eq:spweight}
c_{ij}=\exp\left(\alpha (1-a_{ij})\right),
\end{equation}
where $a_{ij}$ is the neighbor order between individual $i$ and individual $j$ and $\alpha$ is also a tuning parameter to be selected using the modified BIC in \eqref{eq:BIC}.  For example, if $i$ and $j$ are neighbors,  then $a_{ij}  = 1$,  and if $i$ and $j$ are not neighbors, but they share neighbors, then $a_{ij}=2$. If $i$ and $j$ are not neighbors and do not share neighbors, but their neighbors are neighbors, then the neighbor order will be 3. Similarly, all values of $a_{ij}$ can be defined. Figure \ref{fig:neighboor} gives an example of this definition for point 0. For points with ``1", they are neighbors of ``0", for other points, the neighbor orders are also defined. And the neighbor order can be considered as a measure of distance. When individual $i$ and individual $j$ are close in spatial location, the corresponding weight $c_{ij}$ would be large, then two locations will tend to shrink together. More discussions can be found in \cite{wang2019spatial}.  In \cite{wang2019spatial}, they used four candidate values of $\alpha$ to select the best one. Here, a grid of 20 values of $\alpha$ is used, where this grid of values is in a range of 0.05 to 1 with an increment of 0.05. And the best one will be selected based on BIC. In the simulation results when compared to ``FWPe" with equal weights $c_{ij} = 1$ (Tables \ref{tab_KR_setting3}, \ref{tab_sim41} and \ref{tab_sim42}), this range of $\alpha$ can make use of the spatial neighbor information to improve the clustering results. 

\begin{figure}[H]
\centering
\includegraphics[scale=0.6]{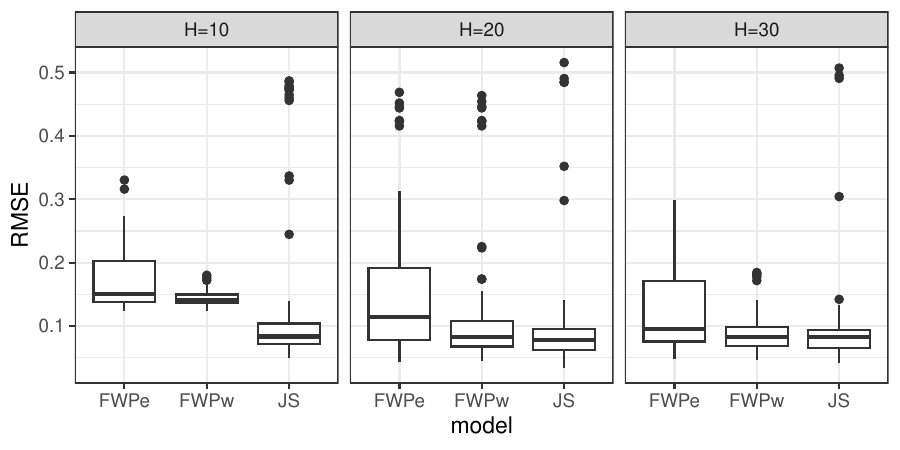}
\caption{RMSE of Scenario 2}
\label{fig:rmse_setting2}
\end{figure}

Since Scenario 1 is the easiest case with more separate mean functions, ``IND" did not perform well compared to other methods because of ignoring the covariance function, thus, this method is not included in other setups. Table \ref{tab_KR_setting2} and Figure \ref{fig:rmse_setting2} show the results of the comparison of different methods. ``FWPe" represents equal weights, that is $c_{ij} = 1$ and ``FWPw" represents pairwise weights in \eqref{eq:spweight}. {\rc Note that, the selection of the number of principal components $P$ does not depend on $c_{ij}$, thus `` FWPe" and ``FWPw" have the same number of components.} From the results, it can be  seen that ``FWPw" performs slightly better than ``FWPe" for  estimating the number of groups,  recovering the group structure and smaller RMSE.  But ``FWPw" is still much better than the other three methods under this simulation. In terms of ARI, ``FWPw" is slightly better than ``JS", especially has a smaller standard deviation. 

\begin{table}[h]
\centering
\caption{ Summary of $\hat{K}$ and average ARI for Scenario 2}
\label{tab_KR_setting2}
\begin{tabular}{cc|ccc}
\hline
& &  $H =10$ & $H=20$ & $H = 30$\\
\hline
\multirow{5}{*}{$\hat{K}$} & FWPe & 3.31(0.56) & 4.09(2.27) & 3.41(0.68) \\ 
  &FWPw & 3.01(0.1) & 3.76(2.38) & 3.1(0.33) \\ 
  &JS & 3.06(0.34) & 3.03(0.22) & 3.02(0.2) \\ 
  &FEM & 2.2(0.47) & 4(0.47) & 4.18(0.69) \\ 
  &HDDC & 4.14(0.93) & 4(0.94) & 4.4(0.8) \\ 
  \hline
  \multirow{6}{*}{ARI}& FWPe & 0.98(0.033) & 0.94(0.167) & 0.98(0.029) \\ 
 & FWPw & 1(0.001) & 0.95(0.17) & 0.999(0.005) \\ 
  &JS & 0.94(0.169) & 0.97(0.112) & 0.98(0.097) \\ 
  &FEM & 0.55(0.034) & 0.45(0.121) & 0.6(0.221) \\ 
  &HDDC & 0.43(0.098) & 0.6(0.155) & 0.66(0.151) \\ 
  &KMA & 0.78(0.123) & 0.81(0.153) & 0.83(0.142) \\ 
\hline 
components &   &2(0) &1.93(0.25) &2(0)  \\
\hline
\end{tabular}
\end{table}

\subsection{Scenario 3}

Several sets of mean functions are considered,  and functions are more similar.

\begin{figure}[H]
\centering
\includegraphics[scale=0.6]{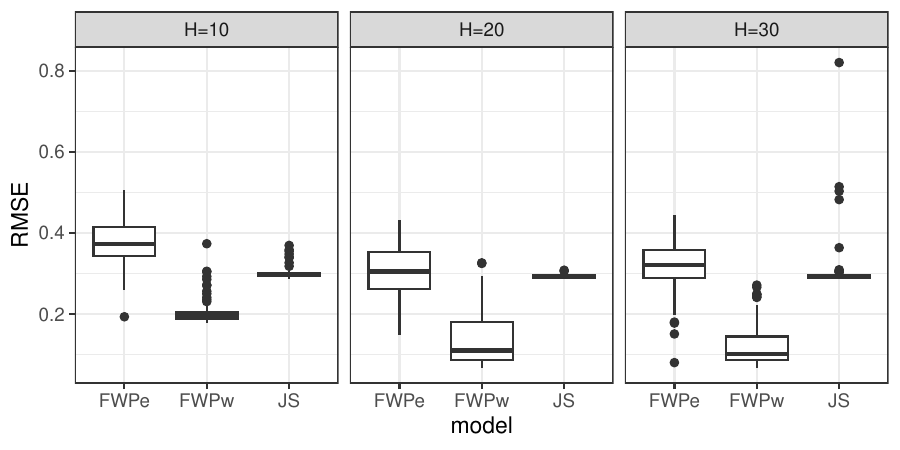}
\caption{RMSE of Scenario 3}
\label{fig:rmse_setting3}
\end{figure}

\paragraph*{Example 1}
Three mean functions are  $\mu_1(t) = \sqrt{2}\sin(4\pi t) + 1$, $\mu_2(t) = \sqrt{2} \sin(4\pi t) + 0.3$ and $\mu_3(t) = 2.5\exp(-25(t-0.25)^2) + 2\exp(-50(t-0.75)^2)$. The group structure in Scenario 2 (Figure \ref{fig:groupsp}) is used.  Table \ref{tab_KR_setting3} and Figure \ref{fig:rmse_setting3} show the results comparing equal weights (FWPe),  pairwise weights (FWPw) and other methods. From the results, it can be shown that the model with pairwise weights (FWPw) performs much better than the model with equal weights,  the JS method and the other three methods in this scenario. The reason is that the three groups are similar and the group structure depends on the location. In the model with pairwise weights, the spatial information is incorporated to improve the estimator's performance.

\begin{table}[h]
\centering
\caption{ Summary of $\hat{K}$ and average ARI for Scenario 3}
\label{tab_KR_setting3}
\begin{tabular}{cc|ccc}
\hline
& &  $H =10$ & $H=20$ & $H = 30$\\
\hline  
  \multirow{5}{*}{$\hat{K}$}  & FWPe & 6.61(1.63) & 5.58(1.34) & 5.93(1.12) \\ 
 & FWPw & 3.19(0.44) & 3.61(0.97) & 3.38(0.65) \\ 
 & JS & 2.35(1.12) & 2(0) & 2.14(0.7) \\ 
 & FEM & 3.41(0.75) & 3.78(0.5) & 3.84(0.63) \\ 
 & HDDC & 4.72(0.51) & 4.52(0.56) & 4.9(0.3) \\ 
  \hline 
   \multirow{6}{*}{ARI}& FWPe & 0.71(0.131) & 0.78(0.103) & 0.78(0.09) \\ 
  &FWPw & 0.99(0.029) & 0.98(0.053) & 0.99(0.023) \\ 
  &JS & 0.58(0.035) & 0.57(0) & 0.55(0.073) \\ 
 & FEM & 0.46(0.061) & 0.42(0.064) & 0.42(0.076) \\ 
 & HDDC & 0.39(0.079) & 0.45(0.091) & 0.47(0.108) \\ 
 & KMA & 0.32(0.089) & 0.31(0.108) & 0.32(0.128) \\ 
 \hline  
components & & 2(0) &2.23(0.43) &2(0)  \\
\hline
\end{tabular}
\end{table}

\begin{figure}[H]
\centering
\includegraphics[scale=0.6]{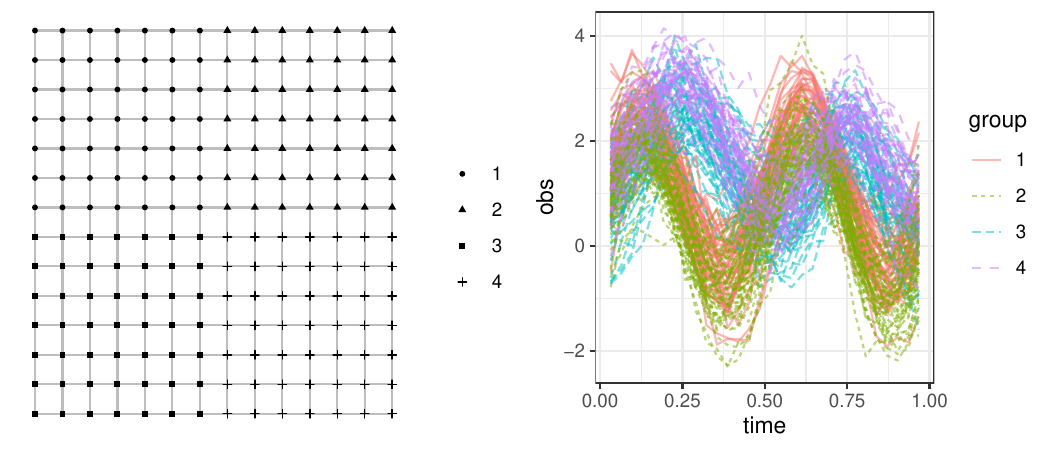}
\caption{The spatial grid and an example of observed curves for $K=4$ }
\label{fig:example4}
\end{figure}

\paragraph*{Example 2} Two examples with $K=4$ are considered to illustrate the performance when there are more groups.  In the first set of mean functions,  four group mean functions are  $\mu_1(t) = \sqrt{2}\sin(4\pi t) + 1,  \mu_2(t) = \sqrt{2}\sin(4\pi t) + 0.3,\mu_3(t) = 2.5\exp(-25(t-0.25)^2) + 2\exp(-50(t-0.75)^2),  \mu_4(t)  = 2.5\exp(-25(t-0.25)^2) + 2\exp(-50(t-0.75)^2) + 0.7$, where group 1 and group 2 are similar,  group 3 and group 4 are similar.  In the second set, $\mu_1(t)$ and $\mu_3(t)$ are the same as those in the first set,  and $ \mu_2(t) = \sqrt{2}\sin(4\pi t) + 0.5, \mu_4(t)  = 2.5\exp(-25(t-0.25)^2) + 2\exp(-50(t-0.75)^2) + 0.5$.  In these two examples,  each group has 49 individuals and the spatial structure is shown in the left figure of Figure \ref{fig:example4}.  The right figure in Figure \ref{fig:example4} shows an example of observed curves when $H = 30$ for the second set of mean functions.  From this figure, we can see that observations from group 1 and group 2 are not well separated, and observations from group 3 and group 4 are not well separated.  


Table \ref{tab_sim41} and Table \ref{tab_sim42} show the summary results based on 100 simulations.  It can be seen that when $H$ is not small,  ``FWPw" can still recover the group structure for both sets of mean functions.   ``FWPe" with equal weights cannot separate groups well.  For JS method, it can be seen that the estimated number of groups is around 2 and the average ARI is small. However, ``FWPw" can recover the group structure well with average high values of ARI. But the estimated number of groups tends to be larger. This could be because some individuals are not clustered into main groups. But when the group difference becomes even smaller, such as less than 0.5, then all methods cannot separate clusters well. 

\begin{table}[h]
\centering
\caption{Summary of results for the first set of mean functions}
\label{tab_sim41}
\begin{tabular}{cc|ccc}
\hline
& &  $H =10$ & $H=20$ & $H = 30$\\
\hline  
  \multirow{5}{*}{$\hat{K}$}  & FWPe & 10.12(1.82) & 9.19(2.3) & 8.16(1.33) \\ 
 & FWPw & 4.14(0.4) & 4.72(2.64) & 4.23(0.62) \\ 
  & JS & 2(0) & 2(0) & 2(0) \\ 
 & FEM & 3.31(0.79) & 3.67(0.53) & 3.6(0.72) \\ 
 & HDDC & 4.78(0.48) & 4.62(0.55) & 4.91(0.29) \\ 
  \hline 
   \multirow{6}{*}{ARI} & FWPe & 0.74(0.091) & 0.79(0.107) & 0.76(0.087) \\ 
  &FWPw & 0.999(0.003) & 0.982(0.098) & 0.996(0.012) \\ 
  & JS & 0.5(0) & 0.5(0) & 0.5(0) \\ 
 & FEM & 0.39(0.077) & 0.38(0.046) & 0.39(0.057) \\ 
 & HDDC & 0.35(0.044) & 0.39(0.051) & 0.41(0.07) \\ 
 & KMA & 0.35(0.072) & 0.4(0.104) & 0.4(0.099) \\
 \hline  
components & & 2(0) &2.06(0.23) &2(0)  \\
\hline
\end{tabular}
\end{table}

\begin{table}[h]
\centering
\caption{ Summary of results for the second set of mean functions}
\label{tab_sim42}
\begin{tabular}{cc|ccc}
\hline
& &  $H =10$ & $H=20$ & $H = 30$\\
\hline  
  \multirow{5}{*}{$\hat{K}$}  & FWPe & 13.6(0.75) & 12.9(2.84) & 17.74(1.75) \\ 
 & FWPw & 4.2(0.49) & 4.07(0.73) & 5.23(1.7) \\ 
 & JS & 2(0) & 2(0) & 2(0) \\ 
 & FEM & 3.3(0.69) & 3.79(0.48) & 3.63(0.69) \\ 
 & HDDC & 4.72(0.51) & 4.52(0.56) & 4.9(0.3) \\ 
  \hline 
   \multirow{6}{*}{ARI}& FWPe & 0.35(0.047) & 0.38(0.056) & 0.37(0.067) \\ 
  &FWPw & 0.99(0.012) & 0.97(0.12) & 0.99(0.022) \\ 
  &JS & 0.5(0) & 0.5(0) & 0.5(0) \\ 
 & FEM & 0.41(0.043) & 0.37(0.045) & 0.38(0.057) \\ 
 & HDDC & 0.34(0.047) & 0.37(0.045) & 0.33(0.028) \\ 
 & KMA & 0.22(0.048) & 0.22(0.055) & 0.23(0.051) \\
 \hline  
components & & 2(0) &2.06(0.23) &2(0)  \\
\hline
\end{tabular}
\end{table}

\subsection{A homogeneous model}

Besides examples of nonhomogeneous models discussed above,  a case with data simulated from a homogeneous model is considered.  The mean function is $ \mu(t) = \sqrt{2} \sin(4\pi t) + 1$.  150 individual curves are simulated.  Equal weights with $c_{ij} = 1$ are used. Here rand index (RI) is reported because ARI will have 0 value when one of the partitions only has one group.  Table \ref{tab_KR_one} shows the summary results for the estimated number of groups $\hat{K}$,  RI and the estimated number of principal components.  It can be seen that when the number of observations in each curve is small, $H=10$,  the proposed method will estimate more groups and less number of components. But when the number of observations increases,  the proposed method can accurately identify the homogeneous structure from data and estimate the number of principal components.

\begin{table}[h]
\caption{Summary of $\hat{K}$ and average RI for a homogeneous model}
\label{tab_KR_one}
\centering
\begin{tabular}{cccc}
  \hline
 & $H=10$ & $H=20$ & $H=30$ \\ 
  \hline 
 $\hat{K}$ & 3.75(2.76) & 1.05(0.26) & 1.04(0.28) \\ 
  RI & 0.605(0.333) & 0.987(0.068) & 0.989(0.077) \\ 
  components & 2(0) & 2.32(0.469) & 2.04(0.197) \\ 
   \hline
\end{tabular}
\end{table}

\subsection{Evaluation of initial values}

Since the optimization problem is non-convex, appropriate initial values would be important. In the literature using ADMM algorithms and penalty functions to find clusters, most of these works used fixed initial values obtained from ridge regression type of models such as \cite{ma2016exploration}, \cite{zhu2018cluster}, \cite{lv2020nonparametric}, 
 \cite{fang2022biclustering}. In \cite{ma2016exploration} and \cite{zhu2021longitudinal}, they assigned the original initial values to $K$ groups to further improve the initial values. In \cite{ren2021gaussian}, they used both EM and the ADMM algorithm and used $k$-means to obtain the initial values. 
In this part, a simulation study is conducted to evaluate the initial values setup. 

For each simulated data set in Scenario 1 with $H=30$, 100 different initial values are constructed based on the proposed initial value with a random noise from a normal distribution with mean 0 and standard deviation 0.5 and 0.1, respectively. We compare the $\hat{l}$ values in \eqref{eq_lik}, BIC values, ARI and RMSE to the oracle estimator, where the oracle estimator is defined when the true cluster structure is known. For each simulated data set, the smallest $\hat{l}$, smallest BIC, smallest RMSE and the largest ARI are computed for 100 initial values generated by random noises. Table \ref{tab_init} shows the average of different measures for 100 simulated data sets with standard deviation in the parenthesis. When the random initial values are quite different (based on standard deviation 0.5) from the original initial values (described in Section \ref{sec_algorithm}), the results based on the original initial values are better. When the random initial values are close to the original initial values (based on standard deviation 0.1), the random initial values can have slightly better $\hat{l}$, BIC and RMSE, but similar ARI. Together with the simulation results in Sections 4.1, 4.2 and 4.3, the proposed initial values can perform pretty well with larger ARI values. But random initial values based on the original initial values are also suggested to have better results. This approach is used in the real data example in Section \ref{sec_realdata}.

\begin{table}[h]
\centering
\caption{Summary of results for different initial values}
\label{tab_init}
\begin{tabular}{l|llll}
  \hline
 & $\hat{l}$ & BIC & RMSE & ARI \\ 
  \hline
 original initial values& -14657 (210)  & -14057 (220)  & 0.046 (0.073)  & 0.99 (0.018)  \\ 
  standard deviation 0.5 & -8975 (793)  & -7498 (448)  & 0.476 (0.026)  & 0.452 (0.06)  \\ 
  standard deviation 0.1 & -15019 (151)  & -13996 (247)  & 0.033 (0.044)  & 0.992 (0.011)  \\ 
   \hline
\end{tabular}
\end{table}

From the simulation study, it can be concluded that, the proposed method is better than the method without considering covariance structure in terms of estimating the number of groups and mean functions. Besides that, when there is a certain spatial structure in the data set, consideration of spatial weights could also improve the results. Especially when the mean functions are close, the proposed method with weighted penalties  is much better than the other methods discussed in this paper.

\section{A real data example}
\label{sec_realdata}

\begin{figure}[H]
\centering
\includegraphics[scale=0.6]{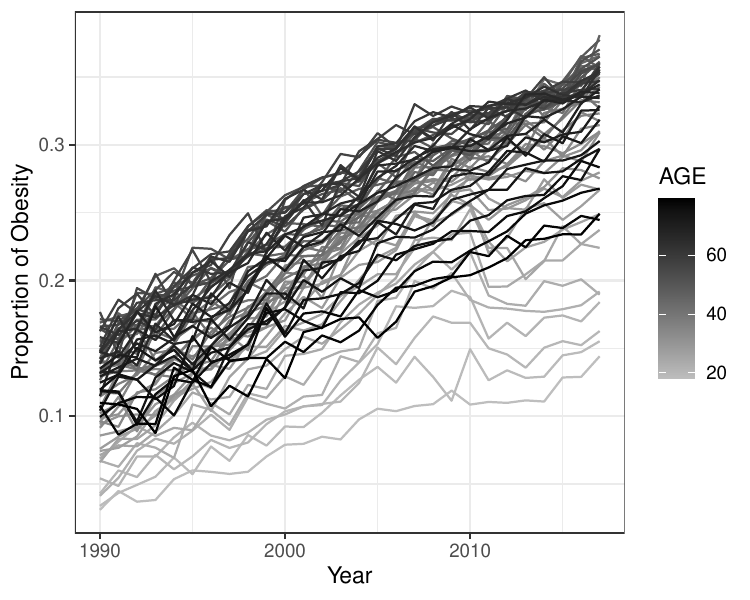}
\caption{Observed curves for different ages over year }
\label{fig:datapattern}
\end{figure}

\begin{figure}[H]
\centering
\includegraphics[scale=0.6]{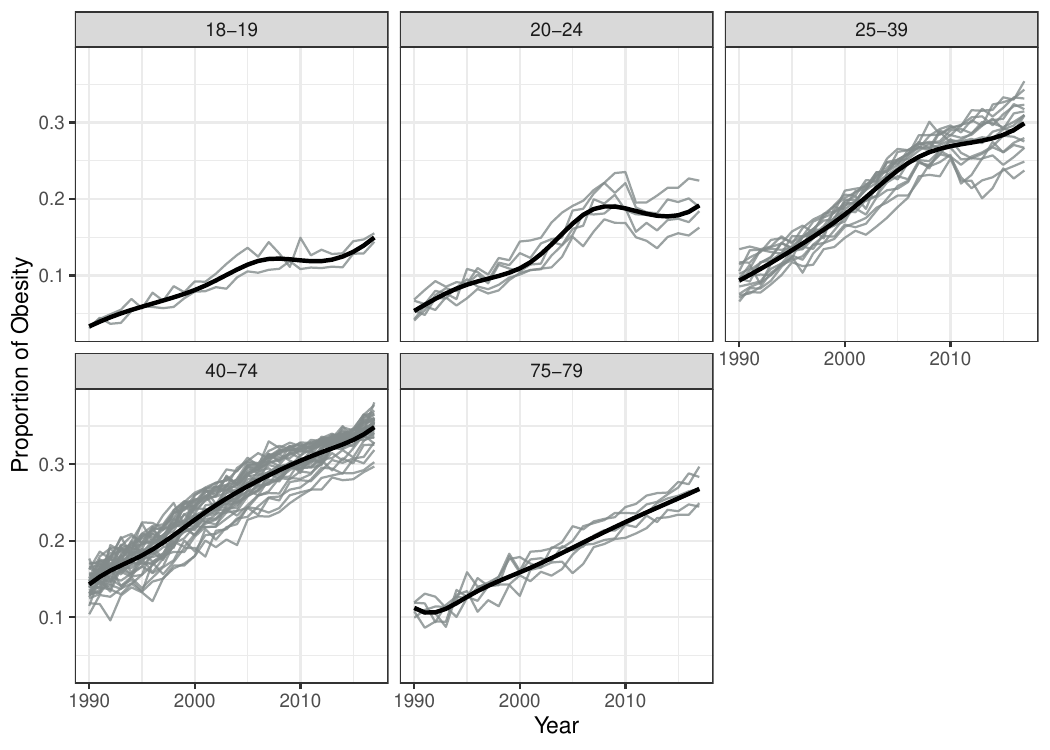}
\caption{Five clusters with the smoothed group curves}
\label{fig:groupfit}
\end{figure}

In this section, the proposed method is applied to a real data set about the obese proportion. The obese proportion data set is aggregated by year and age  based on individual records from the U.S. Department of Health and Human Services,  Center for Disease Control and Prevention (CDC) (\url{https://www.cdc.gov/brfss/annual data/annual data.htm}).  For each age from 18-79, obese proportions from year 1990 to 2017 are obtained.  Figure \ref{fig:datapattern} shows the longitudinal curves for each age. There are some traditional age groups,  such as 20-39, 40-59 and 60+ used by CDC without analyzing the data pattern.    For example,  \cite{hales2017prevalence} used age groups defined by CDC to  analyze prevalence of obesity among adults and youth.  However,  these age groups are somewhat arbitrary,  which don't consider the data trend. \cite{Daawin+Kim+Miljkovic:2019} analyzed this data set with a quadratic trend assumption over time for age curves,  but without considering clusters. \cite{miljkovic2021identifying} considered clustering age curves based on parametric models without considering the covariance structure. 
Instead of using traditional age groups,  a model-based group structure for ages can be found using the proposed method. In order to obtain continuous groups of ages, the following weight is considered, 
$$
c_{ij} = \exp(\alpha(1- \vert i-j\vert)),
$$
where $i$ and $j$ are ages and $\alpha$ is a tuning parameter. This form is also used in \cite{miljkovic2021identifying}.   From the form of the weight, it can be seen that when $\vert i - j\vert = 1$, the weight will take the largest value 1.  If age $i$ and age $j$ are close, the corresponding weight is large and if age $i$ and age $j$ are not close, the corresponding weight is small. 
By using unequal weights,  more shrinkage is put on pairs with closer ages, which will tend to be grouped together. According to the two-step procedure described in Section \ref{sec_simulation}, the number of components is selected first, which is 3. The other two tuning parameters will be selected based on the BIC in \eqref{eq:BIC}. The tuning parameter $\alpha$ is selected in the range from 0 to 1 (with increment 0.05) and $\tau$ is selected in the range from 0.05 to 0.25 (with increment 0.05).  Note that, when $\alpha = 0$,  $c_{ij} = 1$. The combination of $\alpha$ and $\tau$ with the smallest BIC are selected, then the number of clusters and the cluster structure are determined.  Two values of the number of knots, 8 and 9, are used.  Besides the original initial values, 20 random initial values are also used. The result with the smallest BIC is presented. For two different numbers of knots, the result based on 8 knots has smaller BIC value. Besides this, the third component has a very small value of variance, which only contributes about 1\% of the total variation, thus the model with two components is also fitted, which does not change the group structure compared to the three components. Figure \ref{fig:groupfit} shows the estimated group structure and corresponding mean curves based on two components.  All groups are continuous without any discontinuities. The estimated value of $\sigma^2$ is  0.014  and the estimated value of $\lambda_1$ is 0.0485 and $\lambda_2$ is 0.018.

\begin{figure}[H]
\centering
\includegraphics[scale=0.5]{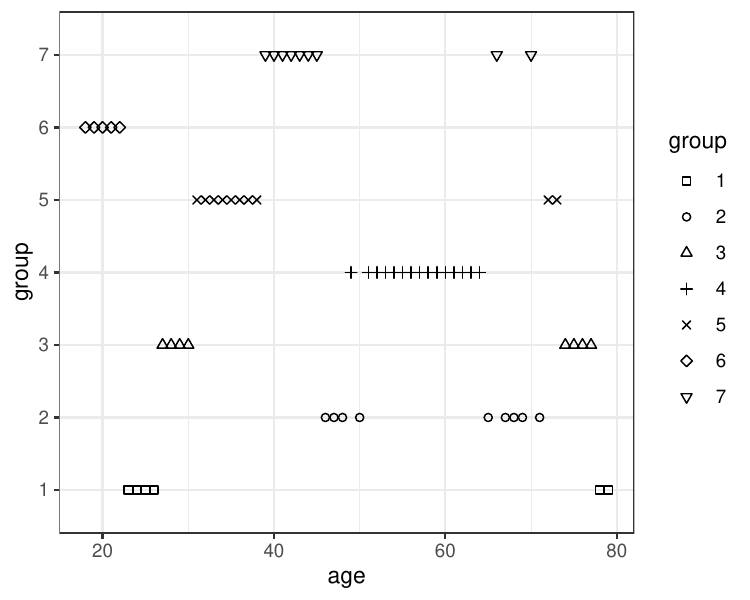}
\caption{Estimated group structures based on the JS method}
\label{fig:groupfit_JS}
\end{figure}

The JS method is also used to analyze this data set. Figure \ref{fig:groupfit_JS} shows the group structure with seven groups selected based on the distortion approach. The estimated group structures based on the JS method are not continuous, which makes the interpretation difficult. 

\section{Summary}
\label{sec_summary}

In this article,  a new method is proposed to find clusters in functional data.  The new method,  FWP,  uses functional principal component analysis to reduce the dimension in the covariance function.  Clusters are identified based on a pairwise concave fusion penalty, which allows different weights in pairwise penalties.  A new algorithm is proposed to solve the constructed optimization problem.   The algorithm combines the EM algorithm and the ADMM algorithm.  The proposed method is compared to some existing methods in the simulation study. The results show that ignoring the covariance structure will reduce the performance in identifying groups. Besides that, the performances of pairwise equal weights (FWPe) and pairwise spatial weights (FWPw) are also considered. The results show that ``spatial weights" performs better than ``equal weights"  and traditional methods when a spatial structure exists and the mean functions are close.

There are some future works that can be considered under this framework. One potential work is to explore the algorithm's performance for functional sparse data and explore the theoretical properties of the proposed estimator. Another potential work is to find clusters of covariance functions together with mean functions. For example, if individual covariance functions are expressed as B-spline basis functions,  corresponding parameters  $\bm{\Theta}_i$ could be grouped for different individuals.  But the algorithm needs to incorporate the constraint of  $\bm{\Theta}_i$ and penalty functions. A new algorithm is needed to solve this problem.  Besides these, the proposed algorithm can be extended to situations to incorporate other covariates with common regression coefficients or clustered regression coefficients, that is $Y_i(t) = \bm{x}_{i}(t)^T\bm{\eta} + X_i(t) + \epsilon_i(t)$, where $\bm{\eta}$ is the vector of common regression coefficients.  Under the proposed framework, pairwise weights can be added to spline coefficients in $X_i(t)$ and estimate $\bm{\eta}$ simultaneously.

\section*{Appendix}

In this appendix,  the EM algorithm with a known group structure is presented.  The EM procedure is similar to the EM algorithm in \cite{james2000principal}, the main difference is that a new design matrix is constructed based on the given group information. 

If the group structure is known, suppose there are $\tilde{K}$ groups and define
$\tilde{\bm{W}}$ be an $n\times\tilde{K}$ matrix with element $w_{ij}$
and $w_{ij}=1$ if $i$ is in the $k$th group. Also define $\bm{W}=\tilde{\bm{W}}\otimes\bm{I}_{q}$
and $\bm{U}=\bm{B}_{0}\bm{W}$. $\left(\tilde{\bm{\alpha}}_{1}^{T},\dots,\tilde{\bm{\alpha}}_{\tilde{K}}^{T}\right)^{T}=\tilde{\bm{\alpha}}=\left(\bm{U}^{T}\bm{U}\right)^{-1}\bm{U}^{T}\bm{Y}$
is the estimate of coefficients for $\tilde{K}$ groups $\bm{\alpha} = (\bm{\alpha}_1^T,\dots, \bm{\alpha}_{\tilde{K}}^T)^T$, which is set as the initial estimate of $\bm{\alpha}$. Thus, $\tilde{\bm{\beta}}_{i}=\tilde{\bm{\alpha}}_{k}$
if $i$ is in the $k$th group. Define 
\[
\bm{C}_{n}=\frac{1}{n}\sum_{i=1}^{n}\left(\bm{\beta}_{i}^{*}-\tilde{\bm{\beta}}_{i}\right)^{T}\left(\bm{\beta}_{i}^{*}-\tilde{\bm{\beta}}_{i}\right), 
\]
where $\bm{\beta}_i^*$ is obtained using the same procedure in Remark \ref{remark_initial}. Then, the eigendecomposition is done for $\bm{C}_{n}=\bm{\Theta}_{0}\bm{\Lambda}_{0}\bm{\Theta}_{0}^{T}$, where $\bm{\Theta}_0$ and $\bm{\Lambda}_0$ are the initial values of  $\bm{\Theta}$ and $\bm{\Lambda}$, respectively. 

Similar to the proposed algorithm, the conditional distribution of $\bm{\xi}_i$ is needed, which has the following forms
\[
\bm{\xi}_{i}\vert\bm{\Omega}\sim N\left(\bm{m}_{i},\bm{V}_{i}\right),
\]
where $\bm{m}_{i}={\rc E\left[\bm{\xi}_{i}\vert\bm{\alpha}, \bm{\Theta}, \bm{\lambda}, \sigma^2 \right]}$ and $\bm{V}_{i}={\rc V\left[\bm{\xi}_{i}\vert \bm{\alpha}, \bm{\Theta}, \bm{\lambda}, \sigma^2 \right]}$ with the following form.
\begin{align*}
\bm{m}_{i}= & {\rc E\left[\bm{\xi}_{i}\vert\bm{\alpha}, \bm{\Theta}, \bm{\lambda}, \sigma^2 \right]}=\left(\bm{\Theta}^{T}\bm{B}_{i}^{T}\bm{B}_{i}\bm{\Theta}+\sigma^{2}\bm{\Lambda}^{-1}\right)^{-1}\bm{\Theta}^{T}\bm{B}_{i}^{T}\left(\bm{Y}_{i}-\bm{U}_{i}\bm{\alpha}\right),\\
\bm{V}_{i}= & {\rc V\left[\bm{\xi}_{i}\vert \bm{\alpha}, \bm{\Theta}, \bm{\lambda}, \sigma^2 \right]}=\left(\frac{1}{\sigma^{2}}\bm{\Theta}^{T}\bm{B}_{i}^{T}\bm{B}_{i}\bm{\Theta}+\bm{\Lambda}^{-1}\right)^{-1}.
\end{align*}
The only difference between the conditional distribution here and the proposed algorithm is that $\bm{\alpha}$ is used instead of $\bm{\beta}$, since the group structure information is given. 

Similarly, $\sigma^2$ is updated by 
\begin{align}
\label{eq:sig20}
\sigma^{2} & =\frac{1}{\sum_{i=1}^{n}n_{i}}\sum_{i=1}^{n}\left(\bm{Y}_{i}-\bm{U}_{i}\bm{\alpha}-\bm{B}_{i}\bm{\Theta}\hat{\bm{m}}_{i}\right)^{T}\left(\bm{Y}_{i}-\bm{U}_{i}\bm{\alpha}-\bm{B}_{i}\bm{\Theta}\hat{\bm{m}}_{i}\right)\nonumber\\
 & +\frac{1}{\sum_{i=1}^{n}n_{i}}\sum_{i=1}^{n}tr\left(\bm{B}_{i}\bm{\Theta}\hat{\bm{V}}_{i}\bm{\Theta}^{T}\bm{B}_{i}^{T}\right).
\end{align}
Also, the same procedure is used to updated $\bm{\Theta}$ and $\bm{\lambda}$ with
\begin{align*}
\tilde{\bm{\theta}}_{j} & =\left(\sum_{i=1}^{n}\bm{B}_{i}^{T}\bm{B}_{i}\left(\hat{m}_{ij}^{2}+\hat{\bm{V}_{i}}\left(j,j\right)\right)\right)^{-1}\\
 & \cdot\sum_{i=1}^{n}\bm{B}_{i}^{T}\left[\left(\bm{Y}_{i}-\bm{U}_{i}\bm{\alpha}\right)\hat{m}_{ij}-\sum_{l\neq j}\bm{B}_{i}\bm{\theta}_{l}\left(\hat{m}_{il}\hat{m}_{ij}+\hat{\bm{V}}_{i}\left(l,j\right)\right)\right].
\end{align*}
Last, $\bm{\alpha}$ is updated as
\[
\tilde{\bm{\alpha}}=\left(\bm{U}^{T}\bm{U}\right)^{-1}\bm{U}^{T}\left(\bm{Y}-\bm{B}_{0}(\bm{I}_{n}\otimes\bm{\Theta})\hat{\bm{m}}\right).
\]

\bibliographystyle{apalike} 
\bibliography{FDA_subgroup}

\end{document}